\documentclass[prd, aps, showpacs,12pt]{revtex4}
\usepackage{latexsym, graphicx}

\def\be{\begin{equation}}
\def\ee{\end{equation}}

\def\te{\end{equation}}
\def\bea{\begin{eqnarray}}

\def\tea{\end{eqnarray}}
\begin{document}

\title{Friction factor for turbulent flow in rough pipes from Heisenberg's closure hypothesis}
\author{Esteban Calzetta$^{1}$  \\
$^{1}${\small CONICET and Departamento de Fisica, FCEN}\\
{\small Universidad de Buenos Aires- Ciudad Universitaria, 1428 Buenos
Aires, Argentina}}
\date{\today }

\begin{abstract}
We show that the main results of the analysis of the friction factor for turbulent pipe flow reported in G. Gioia and P. Chakraborty (GC), Phys. Rev. Lett. 96, 044502 (2006) can be recovered by assuming the Heisenberg closure hypothesis for the turbulent spectrum. This highlights the structural features of the turbulent spectrum underlying GC's analysis.
\end{abstract}

\maketitle
\newpage
\section{Introduction}
The accurate prediction of the friction factor for the turbulent flow of an incompressible fluid through a straight pipe of constant circular section is a matter of huge importance, both practical and fundamental \cite{Sch50,MonYag71,Pop00}. Not surprisingly, it has been the subject of careful measurement \cite{Sch50,McZaSm05}. There is also an array of empirical formulae to predict the friction factor in concrete situations \cite{Cor08,AfzSee07}, culminating with the sheer rendition of Nikuradse's experimental results in analytical form provided by Yang and Joseph \cite{YanJos08}. However, the theoretical link between the phenomenological formulae and the experimental results is weak. In particular, some of the most used empirical formulae, such as Colebrook's, erase most of the structure seen in experiment.

In these circumstances, the derivation of several of the key features of the dependence of the friction factor with respect to Reynolds number for a given pipe roughness from a concrete theoretical model in \cite{GioCha06} (henceforth called GC) is undoubtedly an important step forward. For some background and further developments on the GC model see \cite{GioBom02,GiChBo06,Gol06,MehPou08,GutGol08}. The analysis in GC departs in important aspects from the view of the problem laid down by pioneers such as Prandtl and von Karman \cite{Sch50} and the classic textbook formulation by Landau \cite{LanLif87}. Therefore it is important to understand what are the fundamental elements underlying the success of the GC model. 

Several key features of the Reynolds number dependence of the friction factor, namely the bellies and the Strickler's regime (see GC), can be regained if the GC analysis is combined with Heisenberg's closure hypothesis \cite{Hei48,Cha49,CalGra02} for the turbulent spectrum. The Heisenberg theory is not generally regarded as a realistic depiction of fully developed turbulence \cite{Bat53,Hin75}. Therefore, the fact that the GC analysis works even if the at best qualitatively correct Heisenberg theory is used instead of a (not yet known) exact turbulent spectrum gives us a new perspective on the inner working of the GC model.

This paper is organized as follows. The next section is of a review nature and presents the basic facts and definitions concerning the friction factor and the GC treatment thereof. Section 3 presents the Heisenberg closure hypothesis and derives the GC friction factor for this form of the spectrum. We conclude with a final appraisal of the GC friction factor formula.

\section{Preliminaries}

\subsection{Definitions}

The goal of this Section is just to put together a basic theoretical description of the Reynolds number dependence of the friction factor. Therefore we shall consider only the simplest case of a single phase incompressible fluid moving within a horizontal pipe. The section of the pipe is circular and the radius is $R$. We assume the flow is well developed, meaning that there is a well defined macroscopic velocity $\mathbf{V}$ at every point (we shall use boldface for vector quantities, except for the Reynolds number $\mathbf{Re}$ to be defined below). $\mathbf{V}$ depends only on the radial coordinate $r$, points in the axial direction $x$, and vanishes at the boundary: $
\mathbf{V}=v\left(r\right)\hat{\mathbf{x}}$ and $
v\left(R\right)=0$. 
$\mathbf{V}$ is automatically divergenceless and obeys the momentum balance equations, which, written in cylindrical coordinates, are

\be
\frac{\partial p}{\partial x}=\frac1r\frac{\partial }{\partial r}r\:\tau^{rx}
\label{nsx}
\ee

\be
\frac{\partial p}{\partial r}=\frac1r\frac{\partial }{\partial r}r\:\tau^{rr}
\label{nsr}
\ee
Where $p$ is the pressure and $\tau$ is the stress tensor. 
Since the RHS of eq. \ref{nsr} depends only on $r$, we the pressure drop ${\partial p}/{\partial x}$ must be $r$-independent.
The eq. \ref{nsx} can be integrated (with a boundary condition imposed by regularity)

\be
\tau\equiv\tau^{rx}=\frac{r }{2}\frac{\partial p}{\partial x}\equiv -\tau_0\left(\frac rR\right)
\label{tc}
\te
where $\tau_0$ is the stress at the wall. This means that to know the stress on the wall it is enough to find the stress anywhere, since it obeys a simple scaling law.

Another important quantity is the average flow $Q$. Together with the mass density $\rho$ and the cross surface $A=\pi R^2$ it defines the mean velocity $V$ according to $Q=\rho\:A\:V$. The fluid is characterized by a molecular (dynamic) shear viscosity $\mu$, (kinematic) $\nu=\mu /\rho$. With these quantities we may construct the most important dimensionless number, namely Reynolds'

\be
\mathbf{Re}=\frac{2RV}{\nu}
\label{Reynolds}
\te
These scales allow us to construct an energy density scale $\epsilon=\rho\:V^2$. The Darcy-Weisbach formula introduces the friction factor $f$ from the ratio of $\tau_0$ to $\epsilon$

\be
\tau_0\equiv\frac f8\rho\:V^2
\label{Darcy}
\te
Our main goal is to find a relationship between $f$ and $\mathbf{Re}$.

\subsection{Main flow regimes}
In this subsection we shall describe the main flow regimes and the corresponding empirical formulae. We shall use as reference Nikuradse's experimental results. For practical reasons, we do not mean the actual results, but rather Yang and Joseph's analytical rendering thereof \cite{YanJos08}. 

For a rough pipe the friction factor does not decrease indefinitely with increasing Reynolds number, but rather converges to a finite value $f_{\infty}$. This allows us to define a parameter $\delta_{\infty}\equiv\epsilon R$ from the condition that

\be
\frac1{\sqrt{f_{\infty}}}=-0.868\ln\left[\frac{\epsilon}{7.48}\right]
\label{sixth}
\te
We shall use this parameter to identify the several series of data from the Nikuradse experiment.

\begin{figure}[htp]
\centering
\includegraphics[scale=.8]{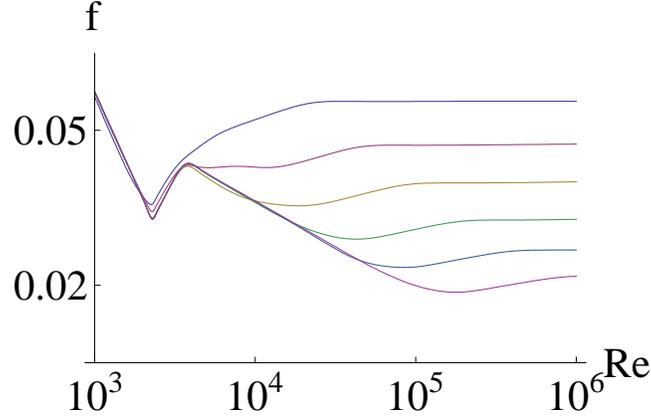}
\caption[f1] {Analytic reconstruction of Nikuradse's data, as given in \cite{YanJos08}, extrapolated to the range up to $\mathbf{Re}=10^6$, actually covered in the Princeton Super Pipe experiment. The six lines correspond, from the bottom up, to $\epsilon^{-1}=507$, $252$, $126$, $60$, $30.6$ and $15$. Both scales are logarithmic.}\label{f1}
\end{figure}

\begin{figure}[htp]
\centering
\includegraphics[scale=.8]{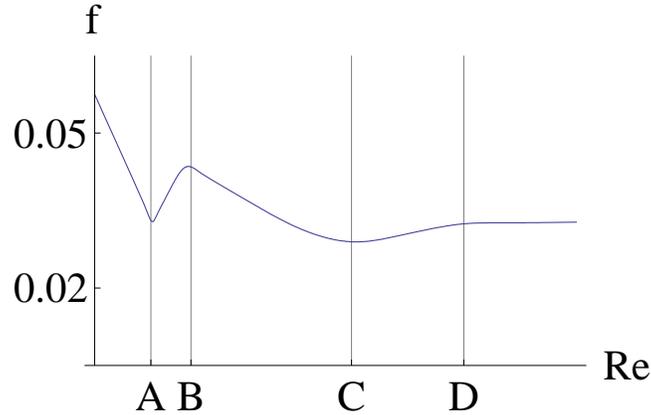}
\caption[f2] {A close up of the curve corresponding to $\epsilon^{-1}=126$. Observe the sharp minimum at $A$, the hump at $B$ and the belly at $C$. To the right of $D$ the friction factor is constant.}\label{f2}
\end{figure}

Fig. \ref{f1} gives an overall impression of the data. Each curve shows a rich structure. For example, let us consider the curve corresponding to $\epsilon^{-1}=126$ (Fig. \ref{f2}). The log-log plot is essentially linear to the left of $A$. This corresponds to the laminar flow; at $A$, we have $\mathbf{Re}\sim 2,000$. Then there is a maximum (the so-called hump) at $B$, corresponding to $\mathbf{Re}\sim 4,000$. The log-log plot is again essentially linear up to the belly at $C$ ($\mathbf{Re}\sim 40,000$). To the right of $C$, the curve approaches its asymptotic value from below. The approach is very fast; beyond the point $D$ ($\mathbf{Re}\sim 200,000$) the friction factor is constant for all practical purposes \cite{YanJos08}

We shall now review two basic regimes in this complex behavior, namely, the laminar regime, to the left of $A$, and the Blasius regime, from the hump $B$ to the belly $C$.

\subsubsection{The laminar regime}
If the flow is laminar,  the bulk velocity profile can be solved exactly and the friction factor results 

\be
f_{\mathrm{laminar}}=\frac{64}{\mathbf{Re}}
\label{laminar}
\te

\begin{figure}[htp]
\centering
\includegraphics[scale=.8]{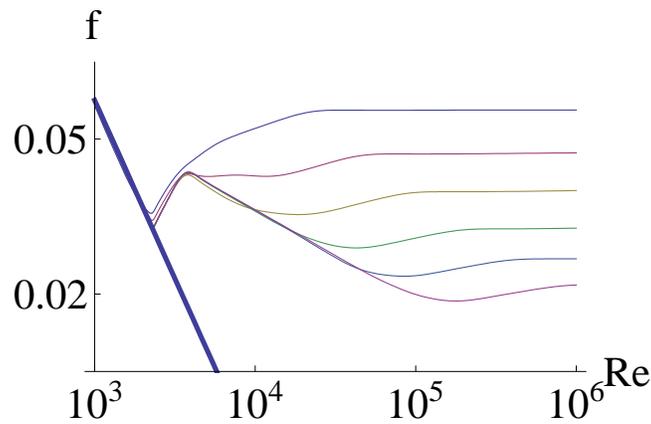}
\caption[f3] {A comparison of the friction factor for laminar flow eq. \ref{laminar} (thick line) against Nikuradse's data as given in fig. \ref{f1}}\label{f3}
\end{figure}
In fig. \ref{f3} we superimpose the plot of the friction factor for laminar flow eq. \ref{laminar} to Nikuradse's data as given in fig. \ref{f1}. We can see that the agreement is outstanding up to Reynolds numbers of a few thousands.

\subsubsection{The Blasius regime}
\begin{figure}[htp]
\centering
\includegraphics[scale=.8]{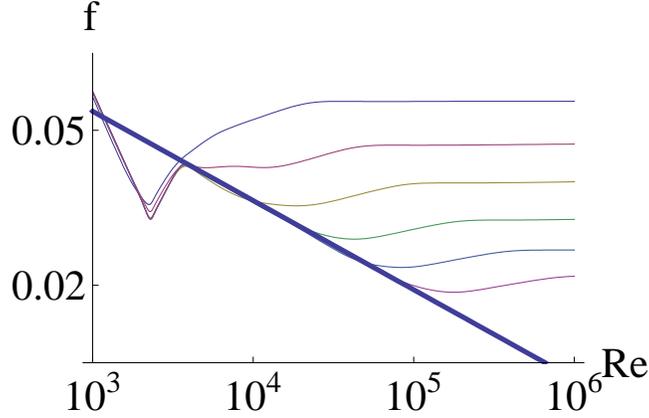}
\caption[f4] {A comparison of Blasius' friction factor  eq. \ref{Blasius} (thick line) against Nikuradse's data as given in fig. \ref{f1}}\label{f4}
\end{figure}

One of the oldest and most accurate empirical formulae for the friction factor is Blasius'

\be
f_{\mathrm{Blasius}}=\frac{0.3164}{\mathbf{Re}^{1/4}}
\label{Blasius}
\te
We plot this expression superimposed to Nikuradse's data in fig. \ref{f4}.

\subsection{The GC model}
In this subsection we shall analyze the proposal by Gioia and Chakraborty (GC) \cite{GioCha06}. Our aim is to contrast it with the  experimental situation as described above.

The basic framework of the GC model is that for high enough Reynolds number the pipe is filled with a well developed turbulent flow which can be accurately described by Kolmogorov's K41 theory. It is therefore characterized by large velocity and length scales $U_{GC}$ and $L_{GC}$. On scales $l$ smaller than $L_{GC}$, the turbulent speed follows Kolmogorov scaling $u_{GC}\left(l\right)=U_{GC}\left(l/L_{GC}\right)^{1/3}$ (except for small corrections to be discussed below). 
GC identify $L_{GC}=R$ as the pipe radius. Concerning the macroscopic flow velocity, GC assume $U_{GC}=\kappa_uV$, where $V$ is the mean flow velocity, and $\kappa_u$ is a $\mathbf{Re}$-\emph{independent} constant (eventually GC choose $\kappa_u=0.036$). 

The detailed mechanisms of momentum transfer between the flow and the wall are rather complex \cite{Town76}. GC assume that the transfer is mainly effected by eddies of size $\delta_{GC}$. These eddies carry a momentum $\rho U_{GC}$  along the wall. The transverse velocity $u_{GC}^0$ is the turbulent speed associated with the scale $\delta_{GC}$

\be
u_{GC}^0\left(\delta_{GC}\right)=U_{GC}\left(\delta_{GC}/L_{GC}\right)^{1/3}
\label{scaling}
\te
The stress at the wall is $\tau_0=\kappa_{\tau}\rho U_{GC}u_{GC}^0$ with some universal constant $\kappa_{\tau}$. Since $\tau_0$ is linear on $u_{GC}^0$, in the Blasius regime $\delta_{GC}\propto\mathbf{Re}^{-3/4}$. This is the same scaling as Kolmogorov's viscous length scale $\eta$. $\eta=R/\mathbf{Re}^{-3/4}$ is the scale at which viscous dissipation matches turbulent energy transport (GC actually interpose another dimensionless constant $b=11.4$ in the definition of $\eta$, we prefer to absorb it into the many more constants to come). Therefore it becomes natural to assume $\delta_{GC}=a\eta$, where $a$ is yet another universal dimensionless constant.  The friction factor reads

\be
f=\frac{8\kappa_{\tau}\kappa_{u}^2a^{1/3}}{\mathbf{Re}^{1/4}}F\left[\mathbf{Re}\right]
\te

where

\be
F\left[\mathbf{Re}\right]=\frac{\mathbf{Re}^{1/4}}{\kappa_{u}a^{1/3}V}u_{GC}^0
\te

At the level of approximation we have stayed so far, $F=1$. If we choose $8\kappa_{\tau}\kappa_{u}^2a^{1/3}=0.3164$, then the model is  built to reproduce Blasius' law for smooth pipes. This can be criticized on the grounds that it does not account for the deviations from the Blasius' Law at very high Reynolds number predicted by the Colebrook equation and apparently verified by the Princeton Superpipe \cite{McZaSm05}. However, these criticisms can be swept away by noting that no physical pipe can be absolutely smooth. So the real issue is how to introduce roughness into the model.

The way to modify the asymptotic behavior of the friction factor to account for pipe roughness is  that the scale $\delta_{GC}$ does not decrease indefinitely with growing Reynolds number, but rather stabilizes at a value $\epsilon_{GC}R$. We therefore find $f_{\infty}\propto\epsilon_{GC}^{1/3}$.

If we identify $\epsilon_{GC}$ with the parameter $\epsilon$ in eq. \ref{sixth}, then this result is the so-called Strickler Law. This approximate scaling was known long before GC's work, and the fact that it is so effortlessly obtained is quite remarkable. 

\subsubsection{The belly and the Blasius regime}

For finite Reynolds number, we may imagine that $\delta_{GC}=\epsilon_{GC}R+a\eta$. This is, the width of the dominant eddies at finite Reynolds number is just the sum of the widths defined by the pipe roughness and by the viscous scale. However, if the stress at the wall is defined by the turbulent velocity at $\delta_{GC}$, and this in turn decays to $\epsilon_{GC}R$ with increasing Reynolds number, then Kolmogorov scaling implies that the friction factor reaches its asymptotic value \emph{from above}, as in the Colebrook equation \cite{Cor08}. To recover the belly in the Nikuradse's data (cfr. figs. \ref{f1} and \ref{f2}) we need that the velocity at an essentially fixed scale $\epsilon_{GC}R$ be an \emph{increasing} function of $\mathbf{Re}$, at least for large enough Reynolds number.

The basic idea is that the mean square velocity at some scale $\delta$ is the sum of contributions from eddies at all scales $k^{-1}<\delta$

\be
u^2\left[\delta\right] =\int_{\delta^{-1}}^{\infty}\:dk\:\mathcal{E}\left[k\right]
\label{global}
\te
$f$ is proportional to $\left(u^2\left[\delta\right]\right)^{1/2}$. If the only dependence on Reynolds number were through $\delta$, then we would obtain

\be
\frac{df}{d\mathbf{Re}}=\frac f{2u^2\left[\delta\right]}\frac{\mathcal{E}\left[\delta^{-1}\right]}{\delta^{2}}\frac{d\delta}{d\mathbf{Re}}<0
\label{posder}
\te
Therefore the friction factor would be a monotonic function of Reynolds number. To get the feature of the belly in the friction factor we have to allow for a direct dependence of the spectrum on Reynolds number, leading to the proper result

\be
\frac{df}{d\mathbf{Re}}=\frac f{2u^2\left[\delta\right]}\left\{\int_{\delta^{-1}}^{\infty}\:dk\:\frac{\partial\mathcal{E}\left[k\right] }{\partial\mathbf{Re}}+\frac{\mathcal{E}\left[\delta^{-1}\right]}{\delta^{2}}\frac{d\delta}{d\mathbf{Re}}\right\}
\label{nonposder}
\te
The derivative does not have a definite sign.

As a matter of fact, this sort of behavior is a prediction of K41 theory for finite Reynolds number. For large but finite Reynolds number, we have

\be
\mathcal{E}\left[k\right]=\frac{3U_{GC}^2}{2R^{2/3}}\:\frac1{k^{5/3}}\:E\left[\beta k\eta\right]
\label{local}
\te
where $E$ is a non increasing function such that $E\left[0\right]=1$ and $E\left[\infty\right]=0$. Thus in the limit of infinite Reynolds number $\eta\mapsto 0$ we recover the one-third scaling law eq. \ref{scaling}. We have added the (yet another) dimensionless constant $\beta$ for latter convenience. Neither $\beta$ nor the form of $E$ more generally are prescribed by K41 theory. This is precisely the point where adoption of the Heisenberg closure hypothesis (or any other, see \cite{L'vPrRu08}) makes a difference.

We observe that inclusion of the $E\left[\beta k\eta\right]$ factor enforces a decay faster than $-5/3$ in the dissipative range. In any case, we now have

\be
\frac{\partial\mathcal{E}\left[k\right] }{\partial\mathbf{Re}}=\frac{3}4\frac{3U_{GC}^2}{2R^{2/3}}\:\frac1{k^{2/3}}\frac{\beta\eta}{\mathbf{Re}}\left(-E'\left[\beta k\eta\right]\right)>0
\te
So the derivative in \ref{nonposder} may be positive in the appropriate range.

We can derive a more explicit result. Using eq. \ref{local} into \ref{global} and calling $x=\beta k\eta$  
we get, after an integration by parts,

\be
u^2\left[\delta\right] =U_{GC}^2\left\{\left(\frac{\delta}R\right)^{2/3}E\left[\frac{\beta\eta}{\delta}\right]-\left(\frac{\beta\eta}R\right)^{2/3}\int_{\frac{\beta\eta}{\delta}}^{\infty}\:\frac{dx}{x^{2/3}}\left(-E'\left[x\right]\right)\:\right\}
\label{fullscaling}
\te
If we use eq. \ref{fullscaling} to evaluate $u$ at the scale  $\delta_{GC}=\epsilon_{GC}R+a\eta$, we clearly see the two asymptotic regimes. For large Reynolds number, $\eta\mapsto 0$, $\delta_{GC}\mapsto\epsilon_{GC}R$ and $u=U_{GC}\epsilon_{GC}^{1/3}$ or, retaining first order corrections

\be
u=U_{GC}\epsilon_{GC}^{1/3}\left\{1-\frac B{\sqrt{\mathbf{Re}}}\right\}
\te
provided $E\left[\beta\eta /\delta\right]=1+o\left(\eta^{2/3}\right)$, where the constant $B$ depends on the shape of the function $E$ as well as on $a$, $\epsilon_{GC}$ and $\beta$

\be
B=\frac12\left(\frac{\beta}{\epsilon_{GC}}\right)^{2/3}\int_{0}^{\infty}\:\frac{dx}{x^{2/3}}\left(-E'\left[x\right]\right)
\te

In the limit where $a\eta\gg\epsilon_{GC}R$ we get instead

\be
u^2\left[a\eta\right] =U_{GC}^2\left(\frac{a\eta}R\right)^{2/3}\left\{E\left[\frac{\beta}a\right]-\left(\frac{\beta}a\right)^{2/3}\int_{\frac{\beta}a}^{\infty}\:\frac{dx}{x^{2/3}}\left(-E'\left[x\right]\right)\:\right\}
\label{fullscaling2}
\te 
and so we recover Blasius' Law with a new constant (unless $\beta /a\ll 1$). Of course this new factor can be easily absorbed in any of the several dimensionless constants at our disposal.

This qualitative success is not easily transformed into a quantitative fit, however. Generally speaking, to get the fast approach to the asymptotic value characteristic of Nikuradse's data, very small values of $\beta$ are preferred. But then the Blasius regime appears in a range of Reynolds numbers much below the experimentally observed. In any case, we have not yet accounted for the hump. We shall defer further quantitative analysis until we incorporate the hump into the GC model.

\subsubsection{The hump}

As we have seen, the behavior of the friction factor for Reynolds numbers over a few thousands, according to GC, is the result of the competition of two opposite processes. On one hand, higher Reynolds numbers mean higher lower limits in the integral eq. \ref{global}, thus bringing the friction factor down. On the other, the integrand in eq. \ref{global}, as defined in eq. \ref{local}, increases pointwise with Reynolds number, thus bringing the friction factor up. The first process dominates in the Blasius regime, the second in the climb up to the asymptotic value.

To set a lower limit to the Blasius region, therefore, the simplest is to cut off the integral in \ref{global}, so that it becomes insensitive to the lower limit if this is low enough. The first process then becomes moot, while the second is still operative, and we get a friction factor which grows with Reynolds number.

We may mention that this second modification of the spectrum is totally outside K41 theory. Also that in a certain way it works too well, since the laminar regime is obliterated.

In summary, GC propose the form of the spectrum (cfr. eq. \ref{local})

\be
\mathcal{E}\left[k\right]=\frac{3U_{GC}^2}{2R^{2/3}}\:\frac1{k^{5/3}}\:E\left[\beta k\eta\right]D\left[ kR\right]
\label{localhump}
\te
where

\be
E\left[x\right]=e^{-x}
\te

\be
D\left[ x\right]=\frac{x^{17/3}}{\left[x^2+\gamma\right]^{17/6}}
\te
and $\gamma$ is a dimensionless constant. Introducing a dimensionless integration variable $x=\beta\:k\eta$ the friction factor may be reduced to the form

\be
f=\frac{C}{\mathbf{Re}^{1/4}}\left\{\int_{h\left[\mathbf{Re}\right]}^{\infty}dx\:\frac{x^4e^{-x}}{\left[x^2+g\left[\mathbf{Re}\right]\right]^{17/6}}\right\}^{1/2}
\label{fullGC}
\te
where

\be
h\left[\mathbf{Re}\right]=\frac{\beta}{a+\epsilon_{GC}\mathbf{Re}^{3/4}}
\te

\be
g\left[\mathbf{Re}\right]=\frac{\gamma\beta^2}{\mathbf{Re}^{3/2}}
\te

\begin{figure}[htp]
\centering
\includegraphics[scale=.8]{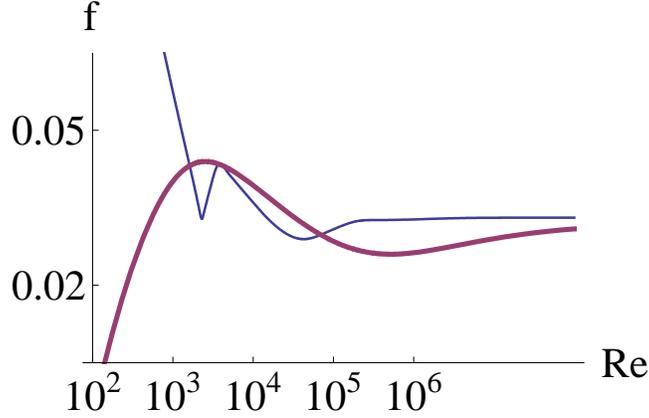}
\caption[f8] {Log-Log plot of the friction factor from eq. \ref{fullGC}, compared to the data for $\epsilon =1/126$. The thick line corresponds to the GC model, the thin line to the data. We have extrapolated the experimental data down to $\mathbf{Re}=10^{2}$ and up to $\mathbf{Re}=10^{10}$ to better appreciate the convergence to the asymptotic values. For this plot, we have chosen $\beta /a=0.5$ and $\gamma\beta^2=10^4$. The values of the constant $C$ and of $\epsilon_{GC}$ were chosen to enforce the Blasius law at the value of $\mathbf{Re}=4000$ and the proper asymptotic limit.}\label{f8}
\end{figure}

In fig. \ref{f8} we show a typical plot of eq. \ref{fullGC}, corresponding to the data for $\epsilon =1/126$. We have extrapolated the experimental data up to $\mathbf{Re}=10^{10}$ to better appreciate the convergence to the asymptotic value. For this plot, we have chosen $\beta /a=0.5$ and $\gamma\beta^2=10^4$. The values of the constant $C$ and of $\epsilon_{GC}$ were chosen to enforce the Blasius law at the value of $\mathbf{Re}=4000$ and the proper asymptotic limit.

This plot represents a near optimal situation; for example, higher values of $\gamma$ erase the features of the curve, while for lower values the hump disappears. Although there is a resemblance to the experimental data, it is not truly quantitatively accurate; in particular, the theoretical prediction is much smoother than the experimental plot.

\section{A GC model based on Heisenberg closure}

In this section we shall use the Heisenberg closure to derive the form of the spectrum in eq. \ref{global}, thereby obtaining an expression for the friction factor with higher predictive power than the original GC proposal. We consider this as a toy model to test whether a turbulent spectrum based on a definite hypothesis regarding the underlying dynamics can make a difference in the accuracy of a model built along the general lines of the GC approach \cite{Tch54}.

\subsection{The Heisenberg closure in the presence of boundaries}
The basic idea of Heisenberg closure is in fully developed turbulent flow the sum of the energy dissipated by viscosity and the energy transported to smaller eddies remains the same in all scales. At a given scale $k_0$, the energy dissipated through viscosity is

\be
Q_{\nu}\left[k_0\right] =2\nu\left\langle \nabla u^2\right\rangle\left[k_0\right] 
\label{qnu}
\te

where

\be
\left\langle \nabla u^2\right\rangle\left[k_0\right] =\int^{k_0}\:dk\:k^2\mathcal{E}\left[k\right]
\te
Heisenberg assumes that the energy transported by turbulence may be written as

\be
Q_{turb}\left[k_0\right] =2\nu_{turb}\left[k_0\right]\left\langle \nabla u^2\right\rangle\left[k_0\right] 
\label{qturb}
\te
where, on dimensional grounds

\be
\nu_{turb}\left[k_0\right]=A\int_{k_0}^{\infty}\:dk\:\sqrt{\frac{\mathcal{E}\left[k\right]}{k^3}}
\label{nuturb}
\te
where $A$ is a dimensionless constant.
The basic assumption is thus 

\be
Q_{\nu}\left[k_0\right] +Q_{turb}\left[k_0\right] =Q_{tot}
\label{qtot}
\te
independent of $k_0$. If $\mathcal{E}\left[k\right]\propto k^{\alpha}$ when $k\mapsto 0$, then $\left\langle \nabla u^2\right\rangle\left[k_0\right] $ scales as $k_0^{\alpha+3}$ while $\nu_{turb}$ scales as $k_0^{\alpha/2-1/2}$, so we must have $\alpha=-5/3$. This means that the kinetic energy increases indefinitely with eddy size, which is incompatible with the presence of the pipe. We therefore adopt the modification suggested by Parker \cite{Par53}, namely, we replace $\left\langle \nabla u^2\right\rangle\left[k_0\right] $ in eqs. \ref{qnu} and \ref{qturb} by

\be
\left\langle \nabla u^2\right\rangle_R\left[k_0\right] =\int^{k_0}\:dk\:\left(k^2+K^2\right)\mathcal{E}\left[k\right]
\te
where $K\approx R^{-1}$. The new term accounts for the increase in dissipation due to eddy deformation. Our power counting argument now gives $\alpha =-1/3$, so that the total kinetic energy in the flow is finite.

The starting point is than the balance equation

\be
\left\{\nu+\nu_{turb}\left[k\right]\right\}2\left\langle \nabla u^2\right\rangle_R\left[k\right]=Q_{tot}
\label{qRtot}
\te
A derivative of eq. \ref{qRtot} gives

\be
\frac{Q_{tot}}{\left\langle \nabla u^2\right\rangle_R}\frac d{dk}\left\langle \nabla u^2\right\rangle_R-\left\langle \nabla u^2\right\rangle_R\frac{2A}{k^{3/2}}\sqrt{\mathcal{E}}=0
\te
but also

\be
\mathcal{E}=\frac1{k^2+K^2}\frac d{dk}\left\langle \nabla u^2\right\rangle_R
\label{simple}
\te
so, if $\xi =k^2$

\be
\frac d{d\xi}\left\langle \nabla u^2\right\rangle_R^{-3}=\frac{-6A}{Q_{tot}^2}\frac 1{\xi^2}\frac 1{\xi +K^2}
\te
we read the boundary condition off eq. \ref{qtot}

\be
\left\langle \nabla u^2\right\rangle_R^{-3}=\frac{8\nu^3}{Q_{tot}^3}+\frac{6A}{K^2Q_{tot}^2}\left\{\frac1{k^2}-\frac1{K^2} \ln\left[1+\frac{K^2}{k^2}\right]\right\}
\te
and then, from eq. \ref{simple}, we get the spectrum from a simple derivative

\be
\mathcal{E}=\frac1{k^3\left(k^2+K^2\right)^2}\frac{4A}{Q_{tot}^2}\left\{\frac{8\nu^3}{Q_{tot}^3}+\frac{6A}{K^2Q_{tot}^2}\left(\frac1{k^2}-\frac1{K^2} \ln\left[1+\frac{K^2}{k^2}\right]\right)\right\}^{-4/3} 
\label{Heisenberg}
\te

We adopt units where $K=R^{-1}=1$. In these units, $Q_{tot}=V^3$ and $\nu =2V/\mathbf{Re}$. Introducing 

\be
K_s\approx\left(\frac{3A}4\right)^{1/4}\frac{\mathbf{Re}^{3/4}}2\equiv\left(\frac{\mathbf{Re}}{\mathbf{Re}}_c\right)^{3/4}
\label{microscale}
\te
then the spectrum can be rewritten as

\be
\mathcal{E}=\mathrm{constant}\;\frac23\frac1{k^3\left(k^2+1\right)^2}\left\{\frac1{K_s^4}+2\left(\frac1{k^2}- \ln\left[1+\frac1{k^2}\right]\right)\right\}^{-4/3} 
\label{Heisenberg2}
\te

To obtain the friction factor, we need to integrate the spectrum from a lower scale $K_0$ up to infinity, where

\be
K_0=\frac{K_s}{\left(\epsilon_{Hei}K_s+a_{Hei}\right)}
\label{khei}
\te
After identifying $K_s=1/\eta$, this lower limit is the same as in the GC approach.

Observe that the spectrum \ref{Heisenberg2} behaves as $k^{-5/3}$ in the inertial range. It displays a faster decay $\mathbf{Re}^{4}k^{-7}$ in the dissipative range. It therefore satisfies the criteria discussed in ref. \cite{GioCha06} and in the previous Section for reproducing the Blasius and Strickler scaling, as well as the belly feature in the friction factor plot.

\begin{figure}[htp]
\centering
\includegraphics[scale=.8]{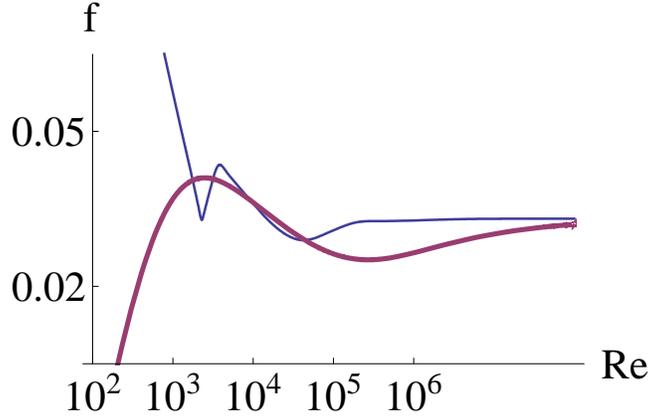}
\caption[f10] {Log-Log plot of the friction factor as computed from the form \ref{Heisenberg2} of the spectrum, compared to the data for $\epsilon =1/126$. The Heisenberg model prediction is the thick line, the experimental values are the thin line. We have extrapolated the experimental data down to $\mathbf{Re}=10^{2}$ and up to $\mathbf{Re}=10^{10}$ to better appreciate the convergence to the asymptotic value. For this plot, we have chosen $a=1.25$, $\epsilon_{Hei}=1/126$ and $\mathbf{Re}_c=1000$. The overall constant is determined by fitting the theoretical value for $f$ at infinitely large Reynolds number to the experimental asymptotic value.}
\label{f10}
\end{figure}

To compare the performance of the model based on Heisenberg closure with the results from the original GC proposal, we shall seek an optimal fit to the Nikuradse data for $\epsilon =1/126$. In all, we have four parameters $\mathbf{Re}_c$, $a_{Hei}$, $\epsilon_{Hei}$ and an overall normalization at our disposal. To reduce parameter space, we shall assume $\epsilon_{Hei}=1/126$ as well. The overall constant is determined by asking for a good fit for very large Reynolds number. Changes in $\mathbf{Re}_c$ induce rigid horizontal shifts in the plot, so $a$ is the parameter which controls the shape of the curve. The best fit is obtained for $a=1.25$, which is physically acceptable. We show the result in fig. \ref{f10}

\begin{figure}[htp]
\centering
\includegraphics[scale=.8]{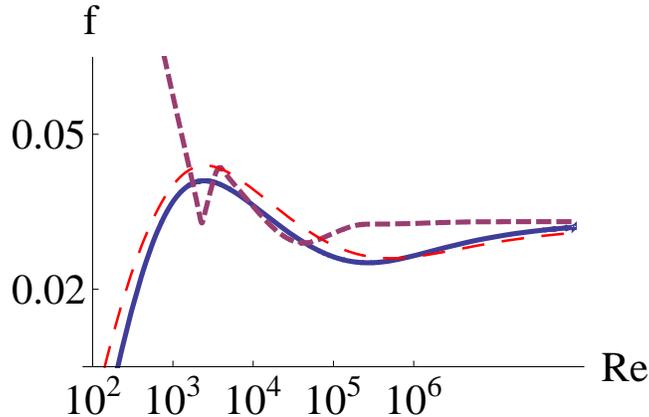}
\caption[f11] {The plots of figs. \ref{f8} and \ref{f10} combined. The experimental data corresponds to the short dashes, the GC model is the long dashes, and the Heisenberg closure is the full line. There is essentially no difference between both theoretical models up to the highest Reynolds numbers.}\label{f11}
\end{figure}

We see that the Heisenberg closure leads to an expression which is as successful as the GC model in describing the Blasius regime and the approach to the asymptotic limit. This is clearly displayed in fig. \ref{f11}, which is simply the superposition of figs. \ref{f8} and \ref{f10}.

\section{Final Remarks}
The GC proposal is striking in that it offers a simple explanation for the ups and downs of the plot of the friction factor against Reynolds number. 

In this note we have combined the main GC postulates with a spectrum derived from Heisenberg closure. The result fits the Nikuradse data as well as the original GC analysis. This is good news for the GC model in that it underlines the fact that the model is built on generic features, rather than detailed dynamical characteristics, of the flow. However, it is also problematic, because Heisenberg closure is not generally regarded as realistic, specially in the dissipative range.

Both in the original and the Heisenberg closure model, moreover, it is clear that the quality of the final fit depends on the careful tuning of the many available parameters.

Our conclusion is that to make progress in understanding the friction factor of turbulent pipe flow along the direction pioneered by Gioia and Chakraborty we do not need to worry about a more accurate spectrum shape, but rather  to provide a solid foundation for the basic assumptions of the GC analysis.

\section*{Acknowledgments}
We wish to thank Jose Lanziani, Salvador Gil and Pablo Mininni for discussions and pointing ref. \cite{Cor08} to us.

This work is supported by University of Buenos Aires, CONICET and ANPCyT.


\begin{thebibliography}{99}
\bibitem{Sch50} H. Schlichting, \emph{Boundary Layer Theory} (McGraw-Hill, New York, 1950).

\bibitem{MonYag71} A. S. Monin and A. M. Yaglom, \emph{Statistical Fluid Mechanics} (MIT Press, Boston, 1971).

\bibitem{Pop00} S. B. Pope, \emph{Turbulent Flows} (Cambridge UP, London, 2000).

\bibitem{McZaSm05} B. J. McKeon, M. V. Zagarola and A. J. Smits, A new friction factor relationship for fully developed pipe flow, J. Fluid Mec. 538, 429 (2005)

\bibitem{Cor08} G. O. Cordero, Improvements in Estimating Pipeline Friction, GIMOR (SPE International-Argentine Petroleum Section-Grupo de interés en modelado y operación de redes y ductos) 7th Annual Meeting, 5 November 2008, Buenos Aires.

\bibitem{AfzSee07} N. Afzal and A. Seena, Alternate scales for turbulent flow in transitional rough pipes: universal log laws, Transactions of the ASME 129, 80 (2007).

\bibitem{YanJos08} B. H. Yang and D. Joseph, Virtual Nikuradse, J. Fluid Mech. (submitted) 

http://www.aem.umn.edu/People/faculty/joseph/PL-correlators/docs-ln/R1-JFM-submissions-virtual-Nikuradse.pdf

\bibitem{GioCha06} G. Gioia and P. Chakraborty, Turbulent friction in rough pipes and the energy spectrum of the phenomenological theory, Phys. Rev. Lett. 96, 044502 (2006).

\bibitem{LanLif87} L. D. Landau and E. M. Lifshitz, \emph{Fluid Mechanics} (Pergamon Press, Oxford, 1987)

\bibitem{GioBom02} G. Gioia and F. A. Bombardelli, Scaling and similarity in rough channel flows, Phys. Rev. Lett. 88, 014501 (2001).

\bibitem{GiChBo06} G. Gioia, P. Chakraborty and F. A. Bombardelli, Rough-pipe flows and the existence of fully developed turbulence, Phys. Fluids 18, 038107 (2006).

\bibitem{Gol06} N. Goldenfeld, Roughness-induced critical phenomena in a turbulent flow, Phys. Rev. Lett 96, 044503 (2006).

\bibitem{MehPou08} M. Mehrafarin and N. Pourtolani, Imtermittency and rough-pipe turbulence, Phys. Rev. E77, 055304(R) (2008)

\bibitem{GutGol08} N. Guttemberg and N. Goldenfeld, The friction factor of two-dimensional rough-pipe turbulent flows, arXiv:0808.1451.

\bibitem{Hei48} W. Heisenberg, On the theory of statistical and isotropic turbulence, Proc. Roy. Soc. A195, 402 (1948).

\bibitem{Cha49} S. Chandrasekhar, On Heisenberg's elementary theory of turbulence, Proc. Roy. Soc. 200, 20 (1949).

\bibitem{CalGra02} E. Calzetta and M. Grana, Reheating and turbulence, Phys. Rev. D65, 063522 (2002).

\bibitem{Bat53} G. K. Batchelor, \emph{The theory of homogeneous turbulence} (Cambridge University Press, Cambridge (England), 1953)

\bibitem{Hin75} J. O. Hinze, \emph{Turbulence} (McGraw-Hill, New York, 1975).

\bibitem{Town76} A. A. Townsend, \emph{The structure of turbulent shear flow}, (Cambridge University Press, London, 1976).

\bibitem{Yag01} A. Yaglom, The century of turbulence theory, in \emph{New trends in turbulence theory}, Les Houches course LXXIV, edited by M. Lesieur, A. Yaglom and F. David (Springer Verlag, Berlin, 2001)

\bibitem{L'vPrRu08} V. L'vov, I. Procaccia and O. Rudenko, Universal model of finite Reynolds number turbulent flow in channels and pipes, Phys. Rev. Lett 100, 054504 (2008).

\bibitem{Tch54} C-M Tchen, Transport processes as foundations of the Heisenberg and Obukhoff theories of turbulence, Phys. Rev. 93, 4 (1954).

\bibitem{Par53} E. Parker, Extension of Heisenber's model of turbulence to critical Reynolds numbers, Phys. Rev. 90, 221 (1953)


\end{thebibliography}
\end{document}